\documentclass[%
aip,
 amsmath,amssymb,%nobibnotes,
 preprint,%
 reprint,%
]{revtex4-1}

\usepackage{graphicx}% Include figure files
\usepackage{dcolumn}% Align table columns on decimal point
\usepackage{bm}% bold math
\usepackage{xcolor}
\usepackage{booktabs}
\usepackage{float}
\newcommand{\todo}[1]{\color{red}TODO: #1 \color{black}}
\usepackage[utf8]{inputenc}
\usepackage[T1]{fontenc}
\usepackage{mathptmx}
\usepackage{lineno}
\usepackage{soul}

\begin{document}

\title[Assessing the Numerical Stability of Physics Models to Equilibrium Variation through Database Comparisons]{Assessing the Numerical Stability of Physics Models to Equilibrium Variation through Database Comparisons}

\author{A. Rothstein}
\affiliation{Princeton University, Princeton, NJ, USA}
\author{V. Ailiani}
\affiliation{Columbia University, New York, NY, USA}
\author{K. Krogen}
\affiliation{Centre College, Danville, KY, USA}
\author{A.O. Nelson}
\affiliation{Columbia University, New York, NY, USA}
\author{X. Sun}
\affiliation{Oak Ridge Affiliate Universities, Oak Ridge, TN, USA}
\affiliation{General Atomics, San Diego, CA, USA}
\author{M.S. Kim}
\affiliation{Princeton University, Princeton, NJ, USA}
\author{W. Boyes}
\affiliation{Oak Ridge Affiliate Universities, Oak Ridge, TN, USA}
\affiliation{General Atomics, San Diego, CA, USA}
\author{N. Logan}
\affiliation{Columbia University, New York, NY, USA}
\author{Z.A. Xing}
\affiliation{General Atomics, San Diego, CA, USA}
\author{E. Kolemen}\email{ekolemen@pppl.gov}
\affiliation{Princeton University, Princeton, NJ, USA}
\affiliation{Princeton Plasma Physics Laboratory, Princeton, NJ, USA}

\begin{abstract}
    High fidelity kinetic equilibria are crucial for tokamak modeling and analysis. Manual workflows for constructing kinetic equilibria are time consuming and subject to user error, motivating development of several automated equilibrium reconstruction tools to provide accurate and consistent reconstructions for downstream physics analysis. These automated tools also provide access to kinetic equilibria at large database scales, which enables the quantification of general uncertainties with sufficient statistics arising from equilibrium reconstruction techniques. In this paper, we compare a large database of DIII-D kinetic equilibria generated manually by physics experts to equilibria from the CAKE and JAKE automated kinetic reconstruction tools, assessing the impact of reconstruction method on equilibrium parameters and resulting magnetohydrodynamic (MHD) stability calculations. We find good agreement among scalar parameters, whereas profile quantities, such as the bootstrap current, show substantial disagreement. We analyze ideal kink and classical tearing stability with DCON and STRIDE respectively, finding that the $\delta W$ calculation is generally more robust than $\Delta^\prime$. We find that in $90\%$ of cases, both $\delta W$ stability classifications are unchanged between the manual expert and CAKE equilibria. 
\end{abstract}
\keywords{kinetic equilibria, DIII-D, MHD, CAKE, tokamak}
\maketitle

\section{Introduction}\label{sec1}

As the global demand for clean and sustainable energy intensifies, magnetic confinement fusion has emerged as a promising long-term solution, with tokamaks providing the most promising avenue for near-term commercialization\cite{buttery_advanced_2021}. Despite significant progress in plasma confinement, stability, and machine integration, several outstanding challenges remain along the tokamak path to fusion energy \cite{bourdelle_integrated_2025,hassanein_potential_2021,petty_diii-d_2019}. While the scope of these challenges are varied, large experimental data sets and predictive modeling capabilities are often used to progress physics understanding and engineering innovations. As devices become more complex, computational tools more sophisticated, and programmatic decisions more consequential, the risks associated with underestimating uncertainties may become increasingly expensive for tokamak analysis, potentially slowing down the realization of commercially viable fusion energy systems. In this light, uncertainty quantification (UQ) is essential not only for improving confidence in theoretical models and simulation tools\cite{} but also for guiding experimental design, optimizing operational regimes, and ensuring the reliability of projections for future devices such as ITER~\cite{hassanein_potential_2021} and DEMO~\cite{zohm_assessment_2013}. 

Equilibrium reconstruction remains an active field of research with unsolved problems related to uncertainty quantification and inconsistencies between reconstruction methods. These unsolved problems stem from equilibrium reconstruction being based on noisy diagnostics\cite{boivin_diii-d_2005}, limited total information about the core plasma, and numerical method difficulties\cite{blum_problems_1990,belli_limitations_2014}. Most advanced analyses on tokamaks start with the reconstruction of a kinetic equilibrium, which orients the plasma profiles, diagnostic locations, and magnetic shape in real space and thus provides essential information to guide downstream physics analyses conducted with advanced physics models \cite{xing_cake_2021,sun_impact_2024,bechtel_accelerated_2022,denk_simultaneous_2025,lao_application_2022,lao_mhd_2005}. The classical approach to this reconstruction involves iteratively improving equilibrium quality with transport, heating and current drive source, and MHD equilibrium codes\cite{lao_mhd_2005,logan_omfit_2018}. More recent approaches have automated this workflow to speed up the reconstruction process, improve consistency\cite{xing_cake_2021}, or even integrate artificial intelligence into the reconstruction framework \cite{lao_application_2022,bechtel_accelerated_2022,sun_impact_2024,madireddy_efit-prime_2024}. A novel approach being pioneered involves a fully integrated Bayesian framework to consistently produce kinetic reconstructions while including UQ in the design of the framework\cite{denk_simultaneous_2025}. 

The results of these downstream analyses are often sensitive to the details of input equilibria. For example, gyrokinetic codes that elucidate relationships between plasma conditions, turbulence, and transport can be highly sensitive to small changes in magnetic structure or profile gradients \cite{poli_experimental_2016,rodrigues_sensitivity_2016,abbate_large-database_2024,snyder_edge_2002}. Time-averaging \cite{nelson_time-dependent_2021,hassan_identifying_2022} and large database studies \cite{nelson_robust_2023,abbate_large-database_2024,turco_causes_2018,bardoczi_onset_2023,groebner_progress_2001,ding_high-density_2024} can be time consuming and expensive for more computationally intensive codes. Limited studies have explored the effect of equilibrium reconstruction on physics analyses, which predominantly study transport\cite{avdeeva_accuracy_2024}. Because of this, it is very important that highly accurate kinetic equilibria be utilized to initiate detailed physics modeling efforts. 

In this paper, we demonstrate variability in equilibrium reconstructions by comparing equilibrium solutions on the DIII-D tokamak generated from automated codes with those created manually by a group of experts. Reconstruction methods are presented in Section~\ref{sec:dataset}, followed by a comparison of scalar equilibrium parameters in Section~\ref{sec:comparison}. Then in Section \ref{sec:MHD}, we explore the changes in results of MHD stability codes for a case study followed by full database analysis before we provide concluding remarks in Section \ref{sec:conc}.

\section{Dataset}\label{sec:dataset}
\begin{figure*}
    \centering
    \includegraphics[width=0.65\linewidth]{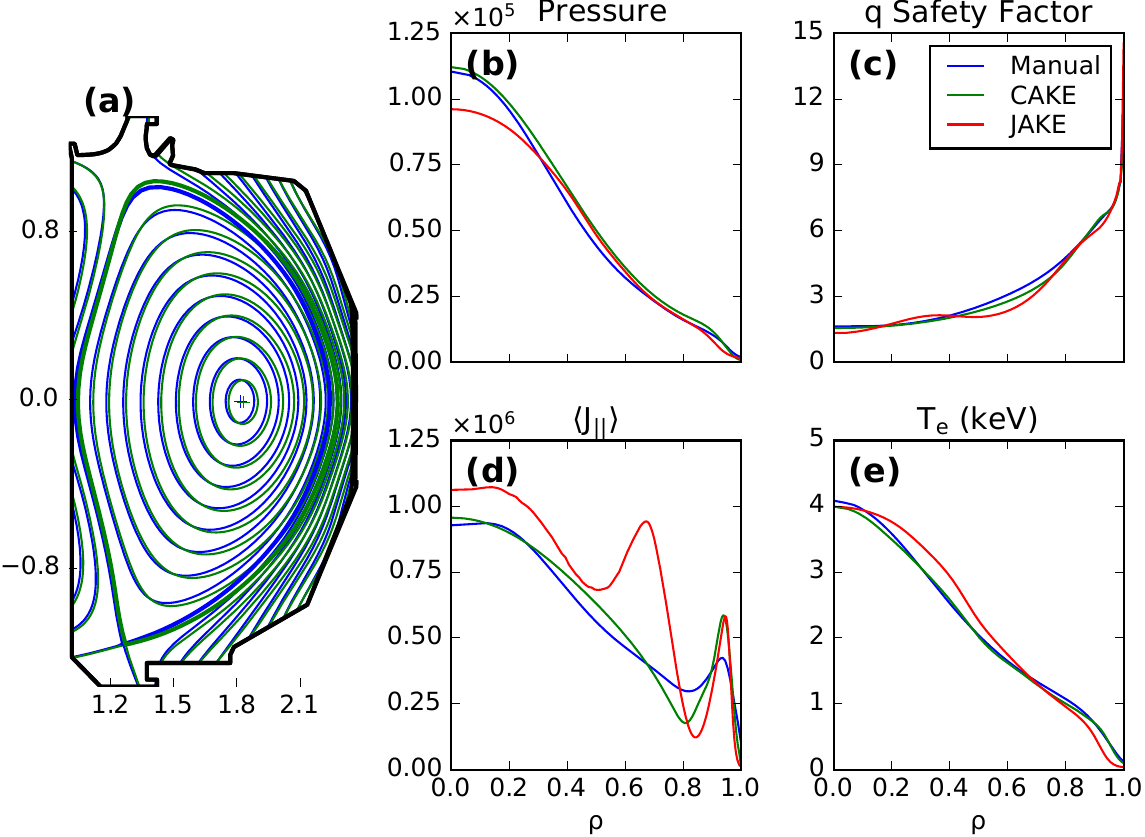}
    \caption{Equilibria parameters for DIII-D shot 180636 at 3900ms. \textbf{(a)} Shows the manual expert (blue) and CAKE 2-dimensional cross sections (green). Note that the JAKE equilibria do not provide a full 2-dimensional cross section of the plasma. \textbf{(b)} Shows the total plasma pressure for manual expert (blue), CAKE (green), and JAKE (red) reconstructions. \textbf{(c-e)} Visualize the manual expert, CAKE, and JAKE equilibria for the safety factor (q), parallel current profile, and electron temperature, respectively.}
    \label{fig:baseProfilesCompare}
\end{figure*}

In this section, we discuss the three types of equilibrium reconstruction used in this study. The first is a set of manual kinetic equilibria constructed by physics experts over several decades at the DIII-D tokamak. These are compared to two automated kinetic equilibrium calculation methods: CAKE \cite{xing_cake_2021} and JAKE \cite{bechtel_accelerated_2022}. Each equilibrium represents a two-dimensional cross section of the plasma, as seen in Figure \ref{fig:baseProfilesCompare}.a, which is characterized by a number of scalar shape parameters defined in Table \ref{tab:parameter_comp} and a 2-dimensional grid of magnetic field vectors and flux functions. Ideal MHD profiles, found by solving the Grad-Shafranov (GS) equation for radial force balance, comprise equilibrium information internal to the separatrix. These are given in Figure \ref{fig:baseProfilesCompare}.b-e. The timeslice shown in Figure \ref{fig:baseProfilesCompare} shows reasonable visual agreement between the expert and CAKE cross sections and equilibrium profiles, whereas the JAKE reconstruction is less physical and in poorer agreement for this example profile. This alone is not necessarily representative of JAKE quality throughout the dataset as both the CAKE and manual expert datasets have timeslices with equally unrealistic profiles for isolated examples.

Notably, all three datasets are subject to challenges in interpretation. At the time of writing, both the CAKE and JAKE codes are under continuous development, which is partially informed by studies such as the one presented below. Further, the manual set of kinetic equilibria, described in more detail in Section~\ref{sec:manual}, is subject to its own uncertainties and thus cannot be taken as a ground truth. This is especially true in light of the fact that different physics experts often arrive at final kinetic equilibrium reconstructions that feature variations depending on the reconstruction techniques employed and the physics goal of each reconstruction. Indeed, resolving user-introduced variations and uncertainties is a central motivation for the automated reconstruction tools \cite{xing_cake_2021}.

\begin{table}
    \centering
    \begin{tabular}{|c|c|c|}
        \hline
        \textbf{Parameter}  & \textbf{Symbol} & \textbf{Range} \\\hline
        Plasma Current & $|I_P|$ (MA) & 0.519-1.99 \\
        Toroidal Magnetic Field & $|B_T|$ (T) & 0.704-2.17 \\
        Bootstrap Current & $|j_{boot}|$ (MA) & 0.258-1.28 \\
        Normalized Pressure & $\beta_N$ & 0.314-4.80 \\
        Plasma Inductance & $l_i$ & 0.441-1.70 \\\hline
        Major Radius & $R_0 (m)$ & 1.66-1.93 \\
        Minor Radius & $a (m) $ & 0.507-0.654 \\
        Triangularity & $\delta$ & -0.496-.856 \\
        Elongation & $\kappa$ & 1.25-2.25 \\
        Plasma Volume &  $Volume(m^3)$  & 12.4-21.1 \\\hline
        Core Safety Factor & $q_0$ & 0.300-9.96 \\
        Safety Factor ($\psi_N=0.95$) & $q_{95}$ & 1.38-13.4 \\
        Core Pressure & $p_{core}$ (kPa) & 14.6-99.7 \\
        Pressure ($\psi_N=0.95$) & $p_{95}$ (kPa) & 0.442-4.99 \\
        Edge Pressure & $p_{edge}$ (kPa) & 0.000-3.87 \\
        \hline
        Core Electron Density & $n_{e,core}$ $(10^{20}/m^3)$ & 0.209-1.56 \\
        Electron Density ($\psi_N=0.95$) & $n_{e,95}$ $(10^{20}/m^3)$ & 0.065-1.11 \\
        Edge Electron Density & $n_{e,edge}$ $(10^{20}/m^3)$ & 0.021-0.546 \\
        Core Electron Temperature & $T_{e,core}$ (keV) & 0.964-9.36 \\
        Electron Temperature ($\psi_N=0.95$) & $T_{e,95}$ (keV) & 0.073-2.23 \\
        Edge Electron Temperature & $T_{e,edge}$ (keV) & 0.009-0.442 \\
         \hline
    \end{tabular}
    \caption{The first column lists names of plasma parameters, while the second column provides their symbolic abbreviation. Core represents a spatial location of $\psi_N=0$, and edge represents a spatial location of $\psi_N=1$, the LCFS. The first horizontal block of parameters represents global parameters, the second block plasma shape parameters, the third block plasma pressure and safety factor related parameters, and the final block electron temperature and density related parameters. The third column provides the full range of each parameter in the Manual dataset. }
    \label{tab:parameter_comp}
\end{table}

\subsection{Manual Expert Equilibria}
\label{sec:manual}
The first equilibrium type considered in this work is a historical dataset of kinetic equilibria created by various physics experts on the DIII-D tokamak with the PyD3D code suite \cite{osborne_enhanced_2015}. Of the many kinetic equilibria previously created for DIII-D, reconstructions from 596 unique shots with 1336 total unique timeslices were selected for inclusion in this study by identifying finalized kinetic equilibria with with Picard iteration errors below $10^{-8}$, which is a typical marker for well-converged cases on DIII-D \cite{xing_cake_2021, lao_reconstruction_1985, logan_omfit_2018}. These timeslices cover over a decade of DIII-D operations from 2012 to 2023 with the corresponding shot range of 150840--195554 and include a representative sampling of the numerous scenarios and plasma characteristics run on DIII-D over that time period. Most of these equilibria were developed and analyzed in support of various peer-reviewed studies. 

To create these reconstructions, various users of the PyD3D code performed an iterative process of mapping profile measurements to the magnetic equilibrium solution in order to simultaneously reduce errors between reconstructed and measured values of magnetic flux and plasma pressure. Though PyD3D provides a standard workflow through which users conduct this process \cite{osborne_enhanced_2015}, individual users may still emphasize fidelity on various parts of the reconstruction (such as core safety factor value or separatrix pressure value) depending on their particular physics goals. For example, some users may have only been interested in the pedestal physics and may have given little thought to the core plasma conditions while others may have done the reverse and focused primarily on the core with lesser focus on the pedestal. The time invested and the resulting equilibrium fidelity also span a range. Each user contributing to this large dataset over the last decade offered different levels of experience and skill with kinetic equilibrium reconstruction, introducing significant variance in the quality of the reconstructions. We have filtered this dataset by Picard iteration error, excluding equilibria whose error > $10^{-8}$. 

Despite these presumed variations, this collection of manual equilibria provides a point of comparison for CAKE and JAKE datasets. These equilibria represent thousands of hours of work done by and deep collective knowledge of a large group of experienced DIII-D users. Given that creating a set of kinetic equilibria involves multiple hours of labor, creating a control dataset of high-quality equilibria this size would be infeasible. This control dataset is the largest dataset of kinetic equilibria available to DIII-D users. Further, any variations in fidelity between individual manually-created kinetic equilibria in this dataset are directly representative of the typical variations introduced into kinetic equilibria through manual reconstruction. The dataset provides a reasonable control set upon which to assess variability from the automated routines.

In summation, despite the variations from the manual kinetic equilibrium procedure, this dataset is the best available and cannot automatically be trusted as the most accurate representation of the plasma as there is still significant variation between individual reconstructions, this dataset is the best available option for describing the historical best practices for kinetic equilibrium reconstruction on the DIII-D tokamak. As such, it is selected and employed in this work as a standard benchmark for various comparisons. While we generally trust this dataset to be highly representative of the real plasmas on DIII-D, on an individual basis it is possible that either the CAKE or JAKE equilibria may be more accurate.

\subsection{CAKE Equilibria}
In place of the manual reconstruction tools, the Consistent Automatic Kinetic Equilibrium, CAKE \cite{xing_cake_2021}, tool provides a consistent reconstruction method that does not vary based on the goal of the study or the experience of the researcher. The CAKE workflow utilizes a similar suite of physics codes to those used in the expert reconstructions. The basic workflow includes: mapping and fitting plasma profiles, generating kinetic constraints with the ONETWO code\cite{pfeiffer_onetwo_nodate}, and generating EFIT and optimizing GS error using the EFIT code\cite{lao_reconstruction_1985}. This optimized plasma reconstruction, EFIT, is used to re-map the plasma profiles, and the process is repeated until adequate convergence is achieved. 

A CAKE equilibrium has been generated for each corresponding expert equilibrium using the default settings for the CAKE workflow. While an exact time to generate the Manual equilibria does not exist, it likely collectively took months to fine tune the dataset while the CAKE workflow was able to calculate the full dataset in less than a week. Further optimizations to the CAKE workflow have been planned on NERSC supercomputing clusters to provide between shot CAKE equilibria. Additionally, CAKE equilibria are automatically generated after DIII-D operation at 100ms time steps and are ideal for initial data exploration without the need to spend significant time manually tuning kinetic equilibria. 

This analysis uses a reconstructions generated by CAKE in July 2023 with the CAKE workflow without MSE constraints and is available on the DIII-D computer clusters. Since the CAKE workflow is actively being updated and improved, a newer CAKE database may yield slightly different results. While a further analysis on the absolute accuracy of CAKE is outside the scope of this work, this iteration of CAKE results provides a platform to analyze numerical stability of MHD codes with respect to different equilibrium reconstruction tools. 

\subsection{CAKE Variations}\label{sec:cakeVar}
While most settings in CAKE have been generalized to work well in any given plasma, there are still options a user can manipulate to influence the equilibrium reconstruction. Due to the combination of multiple reconstruction code submodules, the impact of any individual setting can be hard to predict. For example, small changes in knot locations can have a negligible or significant effect on the resulting equilibrium, which changes on a case to case basis. 

\begin{figure}
    \centering
    \includegraphics[width=\linewidth]{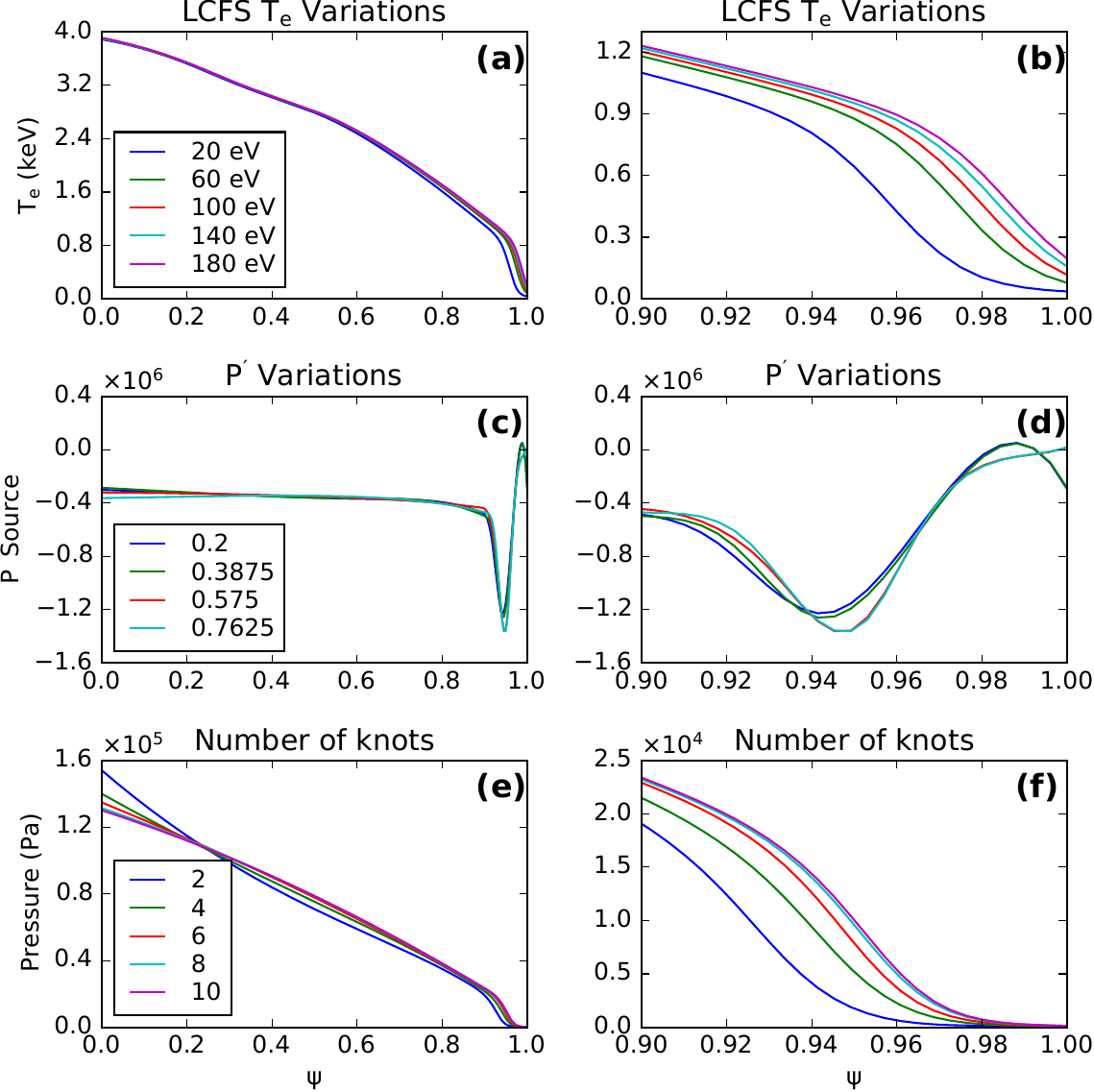}
    \caption{CAKE variations for DIII-D shot 189652 at time 3100ms. The horizontal axis is the normalized poloidal flux $\Psi$ (also reffered to as $\Psi_N$) where the left column is the full range of $\Psi=[0,1]$ and the right column is the pedestal region of $\Psi=[0.9,1]$. \textbf{(a)} and \textbf{(b)} show the effect of varying $T_e$ at the LCFS where \textbf{(b)} is a zoom in on the edge to better see the effect on the pedestal. The value of $T_e$ at the LCFS is given in the legend of \textbf{(a)}. \textbf{(c)} and \textbf{(d)} show the effect of varying the spline knot smoothing on the $P^\prime$ profile, with the scanned smoothing values given in the legend. \textbf{(e)} and \textbf{(f)} show the effect of the number of splining knots on the pressure profile. }
    \label{fig:cakeVariationsComp}
\end{figure}

To quantify the effects of these settings, we have scanned various user settings over our original dataset of CAKE equilibria. Here we present a single base equilibrium with changed CAKE settings while later we will look at a large database scale. The effect of three of these setting variations is shown in Figure \ref{fig:cakeVariationsComp}. The full impact of changing these settings is discussed below in Section \ref{sec:cakeVarComp}. The details regarding CAKE is available in a previous publication\cite{xing_cake_2021}, and here we summarize the CAKE settings examined: 
\begin{itemize}
    \item \textbf{LCFS Temp}: The last closed flux surface electron temperature. The default value for this is typically around $80$eV and we scan this setting from 20-180eV stepping every 40eV. By default, CAKE determines the LCFS surface dynamically by using a modified tanh fit of the $T_e$ profile, but it can be set manually such as for this variation analysis.
    \item \textbf{CER Channels}: The selection of CER channels to be used in the fitting process. Options here involved using all possible channels, using the reduced set of channels that are typically available, or to further filter the channel selection depending on the beam voltage used. In D3D, it is observed that under a certain threshold of neutral beam voltage (70kV), the $C^{6+}$ density measured by corresponding CER channels is systematically lower than the truth. The default is to use the channel filtering that produces results in the correct density measurement.
    \item \textbf{GS Error}: This setting allows additional optimizations to improve the numerical error related to EFIT's iteration scheme. CAKE aims to produce equilibria with iteration error lower than $1e-8$ which have a much higher chance for accurate MHD calculations. Without this further optimization, slightly less than half of CAKE's output slices would reach this error target, where as with this additional optimization, more than 90\% of results would reach this target. While CAKE database is built with additional error minimization enabled, it is not enabled by default during interactive user runs.
    \item \textbf{Spline Uncertainty}: %I am varying CAKE['SETTINGS']['PHYSICS']['profile_fitting']['smoothing_spline_min_psin_uncert}][all_variables}] from 0.001-0.01, which I believe is the allowed uncertainty value that is set for all of the spline smoothing parameters. 
    % Unfortunatly there seems to have been a miscommication between me to Oak to you about what this setting was, but I don't think it impact this paper particularly. --Anthony
    The minimum uncertainty assigned to the measurement location in $\psi_n$, as in the uncertainty in x assumed for the CAKE fitting scheme. CAKE uses an Monte Carlo fitting process to account for position uncertainty. The default values are between 0.005 and 0.01 depending on the diagnostic.
    \item \textbf{Pprime}: 
    % The setting is OMFIT['CAKE_autorun']['SETTINGS']['PHYSICS']['EFIT']['psi_core_pprime']
    Minimum value for spline knot location in the $P^\prime$ profile. The default limit is $0$, meaning unconstrained, and we vary this parameter from $0.0$ to $0.95$. 
    \item \textbf{Ffprim}:
    % Similar to above, this looks like OMFIT['CAKE_autorun']['SETTINGS']['PHYSICS']['EFIT']['psi_core_ffrime']
    Minimum value for the spline know location in the $FF^\prime$ profile, where $F=2\pi R\frac{B_\phi}{\mu_0}$. The default location is $0$, meaning unconstrained, and we vary this from $00$ to $0.95$. 
    \item \textbf{Nffprim}:
    The number of spline smoothing knots in $P^\prime$ and $FF^\prime$ profiles. The default value is $6$ and we vary this from $2$ to $10$.
\end{itemize}

Some of the effects of manipulating the CAKE settings are immediately obvious. In Figure \ref{fig:cakeVariationsComp}.b, the increasing $T_e$ at the LCFS has a clear effect on raising the $T_e$ pedestal without a significant impact on the core $T_e$ profile seen in Figure \ref{fig:cakeVariationsComp}.a. While there is not a clear trend in the effect of changing smoothing knot locations, it has a definitive effect on both the core of the $P^\prime$ profile as well as the pedestal as seen in Figure \ref{fig:cakeVariationsComp}.c-d. Finally, in Figure \ref{fig:cakeVariationsComp}.f, increasing the number of knots appears to increase the pedestal pressure, while there is some minor effect on the core pressure when there are a very low number of splining knots.

\subsection{JAKE Equilibria}
This database corresponds to a series of equilibrium files generated using an automated version of the kineticEFITtime\cite{meneghini_integrated_2015} module found in OMFIT\cite{meneghini_integrated_2015,meneghini_integrated_2013}. This modified, automated workflow used the module to perform kinetic equilibrium reconstruction using the default settings from that module for the DIII-D 2019 experimental run campaign. Additional details of the procedure can be found in \cite{sun_impact_2024}. The discharge range from this campaign spans shot 178544--180916 and, when accounting for the overlap between these discharges and those in the expert database, there are 15 total shots using 29 distinct equilibria for the database comparison.

This automated kineticEFITtime procedure is less robust than the CAKE method, whose development timeline is longer and whose settings and filters are greater in number. This automated procedure intended to quickly generate a large database of kinetic equilibrium reconstructions that could be used as a sample database to perform machine learning studies. In order to differentiate the procedure from CAKE, the colloquial term we chose for this database was ``JAKE.''

While the JAKE dataset is comprised of lower fidelity kinetic reconstructions, it stands as the only other automated kinetic equilibrium reconstruction tool. Ideally, our database comparison would include more equilibria datasets to provide the most data on kinetic equilibria reconstruction methods, however, the Manual, CAKE, and JAKE equilibria are the most comprehensive DIII-D equilibrium databases currently publicly available. As future automated reconstruction workflows are developed, validation and benchmarking should be performed against these datasets to estimate workflow consistency.

\section{Comparing Equilibrium Parameters}\label{sec:comparison}
In this section, we compare the scalar EFIT parameters among datasets. All quantities are considered with percent differences to prevent variations in discharge parameter values e.g. $I_P$ from biasing our findings. We consider equilibria within $\pm50$ms to be a reasonable pairing because this is faster than resistive timescales while being on the same order as heat diffusion timescales.  

\subsection{Profile Comparison}\label{sec:profComp}

\begin{figure*}
    \centering
    \includegraphics[width=\linewidth]{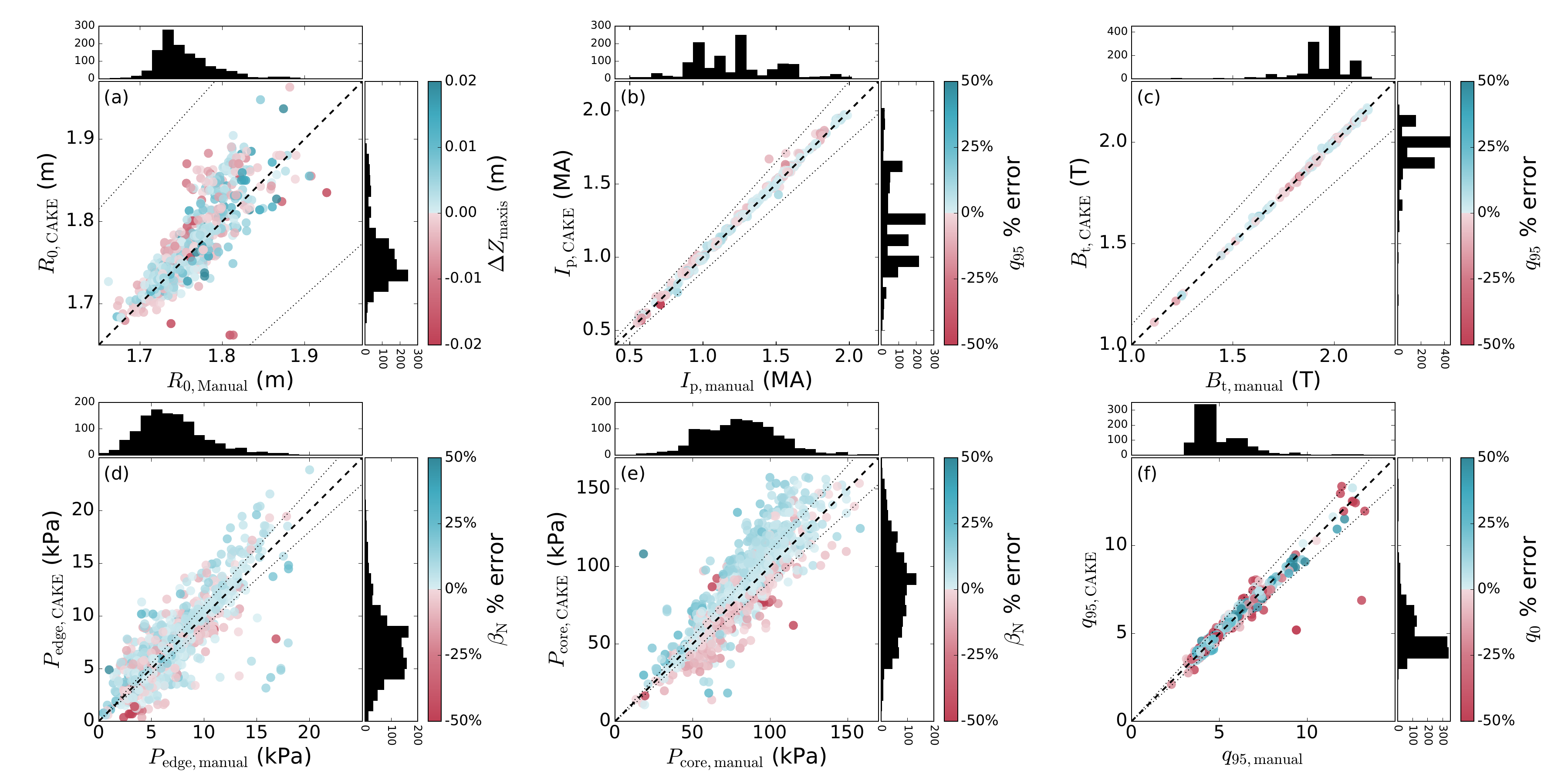}
    \caption{Plotted are equilibrium scalar parameters from manual EFITs (horizontal axes) versus CAKEs (vertical axes). \textbf{(a)} plots major radius $R_0$. The color contour $Z_{axis}$ illustrates the level of disagreement and the histograms above and below show the distribution of the Manual and CAKE datasets, respectively. Plots \textbf{(b)-(f)} follow the same layout for plasma current, toroidal magnetic field, edge plasma pressure, core plasma pressure, and safety factor at $\psi_N=0.95$, respectively.}
    \label{fig:scalarPlots}
\end{figure*}

We visualize the comparison of scalar parameters, like plasma current and core pressure, in Figure \ref{fig:scalarPlots}. Good agreement follows a diagonal line (i.e. $R^2=1$), and disagreement is represented by spread from this line (i.e. $R^2<1$). Good agreement can be seen in plasma current in Figure \ref{fig:scalarPlots}.b, toroidal magnetic field in Figure \ref{fig:scalarPlots}.c, and $q_{95}$ in Figure \ref{fig:scalarPlots}.f. There is significantly more spread in edge pressure in Figure \ref{fig:scalarPlots}.d and core pressure in Figure \ref{fig:scalarPlots}.e, while major radius in Figure \ref{fig:scalarPlots}.a falls somewhere in between. 

\begin{figure}
    \centering
    \includegraphics[width=0.8\linewidth]{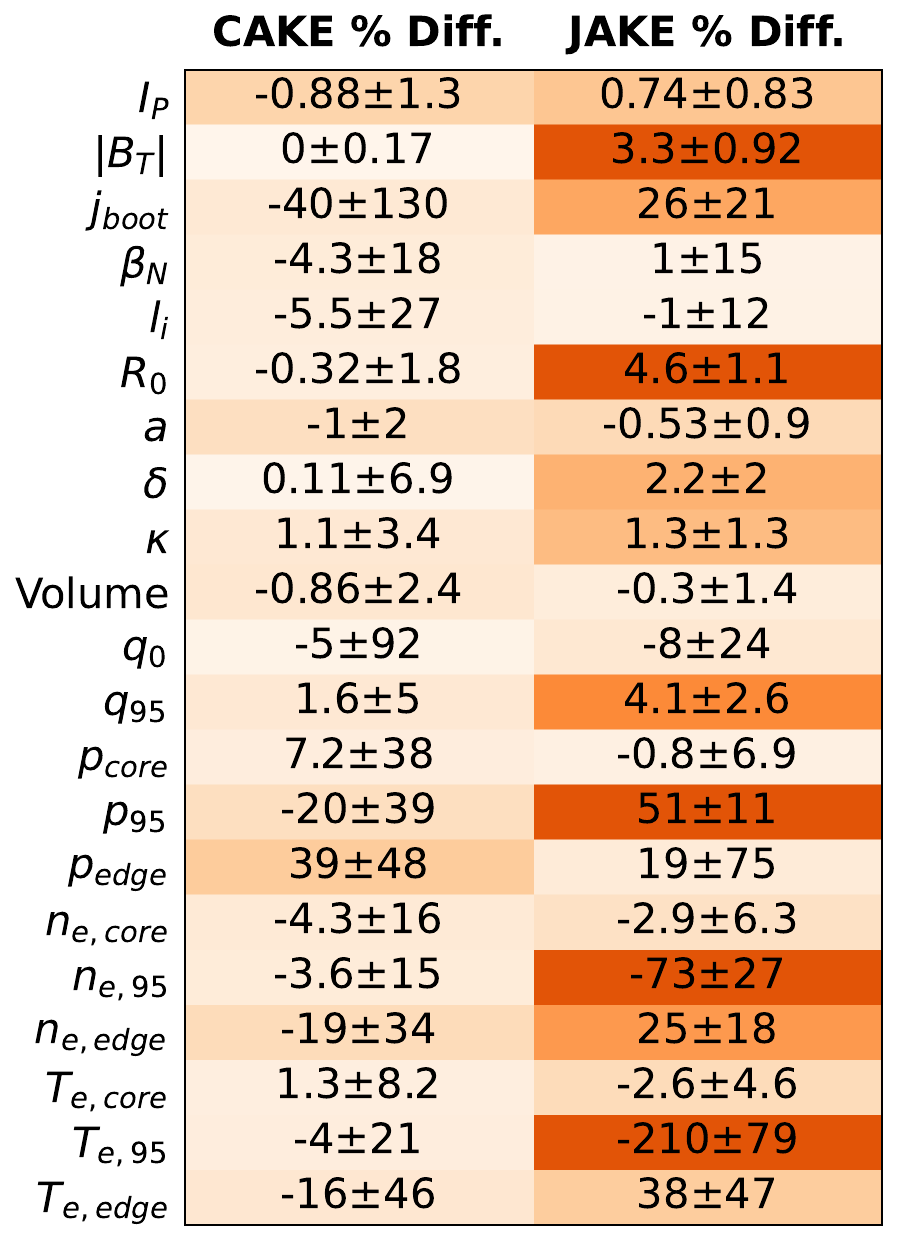}
    \caption{Percent difference of CAKE (left column) and JAKE (right column) compared to the Manual equilibrium dataset. Each value is reported as mean$\pm$standard deviation of the percent differences. Color intensity is proportional to the mean divided by standard deviation. }
    \label{fig:mainComp}
\end{figure}

There was an overlap of 1194 equilibria between the CAKE and Manual datasets and a significantly smaller overlap of 29 equilibria between the JAKE and Manual datasets. Because there were significantly fewer JAKEs that overlapped with the Manual dataset, we will also compare JAKEs to a larger dataset of CAKEs later. To quantify variance in these scalar profiles, we move from raw values to percent differences to normalize all quantities. We calculate the percent difference of the CAKE and JAKE profiles from each Manual equilibria and report the Mean$\pm$Standard Deviation of percent differences for each parameter. We calculate percent differences as $(Manual-CAKE)/Manual$ and $(Manual-JAKE)/Manual$. This metric is reported in Figure \ref{fig:baseProfilesCompare} for each scalar parameter. We color each parameter so the intensity is proportional to the mean percent difference divided by the standard deviation. The effect is such that when the reconstruction methods agree, the mean is close to $0$ and the color is lighter. When the equilibria disagree and the mean percent difference is comparatively further from $0$, the intensity is brighter. 

The first observed fact is there is much more deviation in the JAKE profiles than the CAKE profiles compared to Manual equilibria, as evidenced by the JAKE column having stronger color intensity. Part of this may be attributed to the small sample size, but a likely cause is the lower fidelity in the JAKE workflow, such as the omission of shifting Thomson diagnostic measurements and other automated copmutation choices. Further discussion on the difference will be continued in Section \ref{sec:jakeComp}. 

When comparing the reconstructions, the results generally agree with intuition of kinetic equilibrium construction. Shape parameters such as $R_0$, $a$, $\delta$, and $\kappa$ agree well as they typically are not strongly affected by kinetic constraints. Harder to fit quantities like edge pressure, $q_0$, and $|j_{boot}|$ have high mean percent differences, but even higher standard deviations in the percent differences. This represents high variance between the Manual reconstructions and CAKE or JAKE. The electron temperature and density profile parameters fall in the middle, with more spread than shape parameters but not quite as much as the pressure and safety factor profile parameters, explainable as the diagnostic measurements can be shifted due to separatrix misalignment.

\subsection{Automated Reconstruction Comparison}\label{sec:jakeComp}

\begin{figure}
    \centering
    \includegraphics[width=0.5\linewidth]{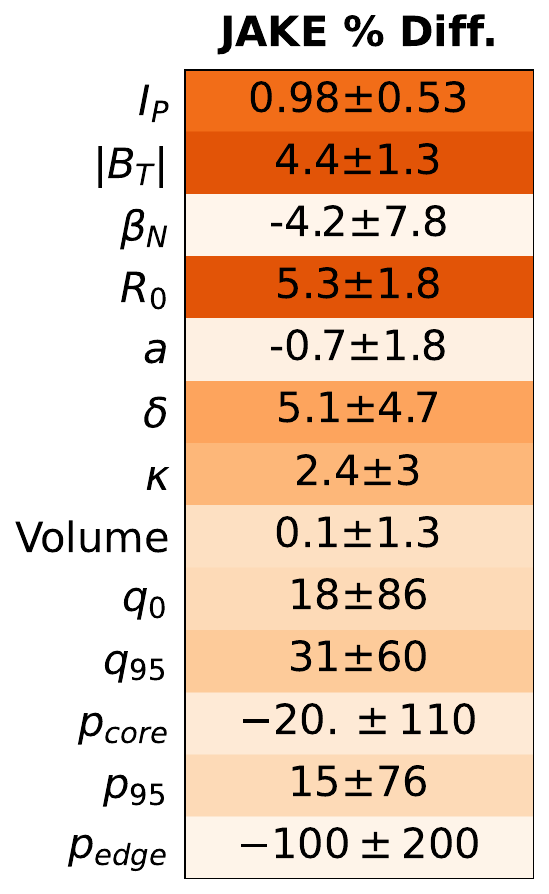}
    \caption{Mean$\pm$standard deviation of scalar parameters for 882 equilibrium pairs of JAKE and CAKE equilibria. Each percent difference is calculated as $100\times(CAKE-JAKE)/CAKE$. Color intensity is given by absolute value of the mean percent difference divided by standard deviation for each parameter. }
    \label{fig:jakeComp}
\end{figure}

To more robustly quantify JAKE equilibria, we look at a
different subset of equilibria of the overlaps between JAKE and CAKE. There are $882$ equilibrium pairs, and the percent differences and standard deviations are given in Figure \ref{fig:jakeComp}. The color intensity of each row is the mean percent difference divided by the standard deviation, and values with a greater color intensity are comparatively further from a mean of $0\%$ than rows with lower color intensities. Taking a closer look, we find that JAKEs generally have minor inconsistencies on some of the global parameters such as $B_T$ and $R_0$. While the discrepancy in $R_0$ location may be attributed to the lack of Thomson measurement shifts, as seen by the LCFS electron temperature being important for $R_0$ in CAKE variations in Section \ref{sec:cakeVarComp}, it is at this moment unclear if there is some other underlying cause for the major axis difference. While the remaining parameters for shape, pressure, and safety factor profiles have some variance, they are generally within the bounds of a single standard deviation.

\subsection{CAKE Setting Variants}\label{sec:cakeVarComp}

\begin{figure*}
    \centering
    \includegraphics[width=\linewidth]{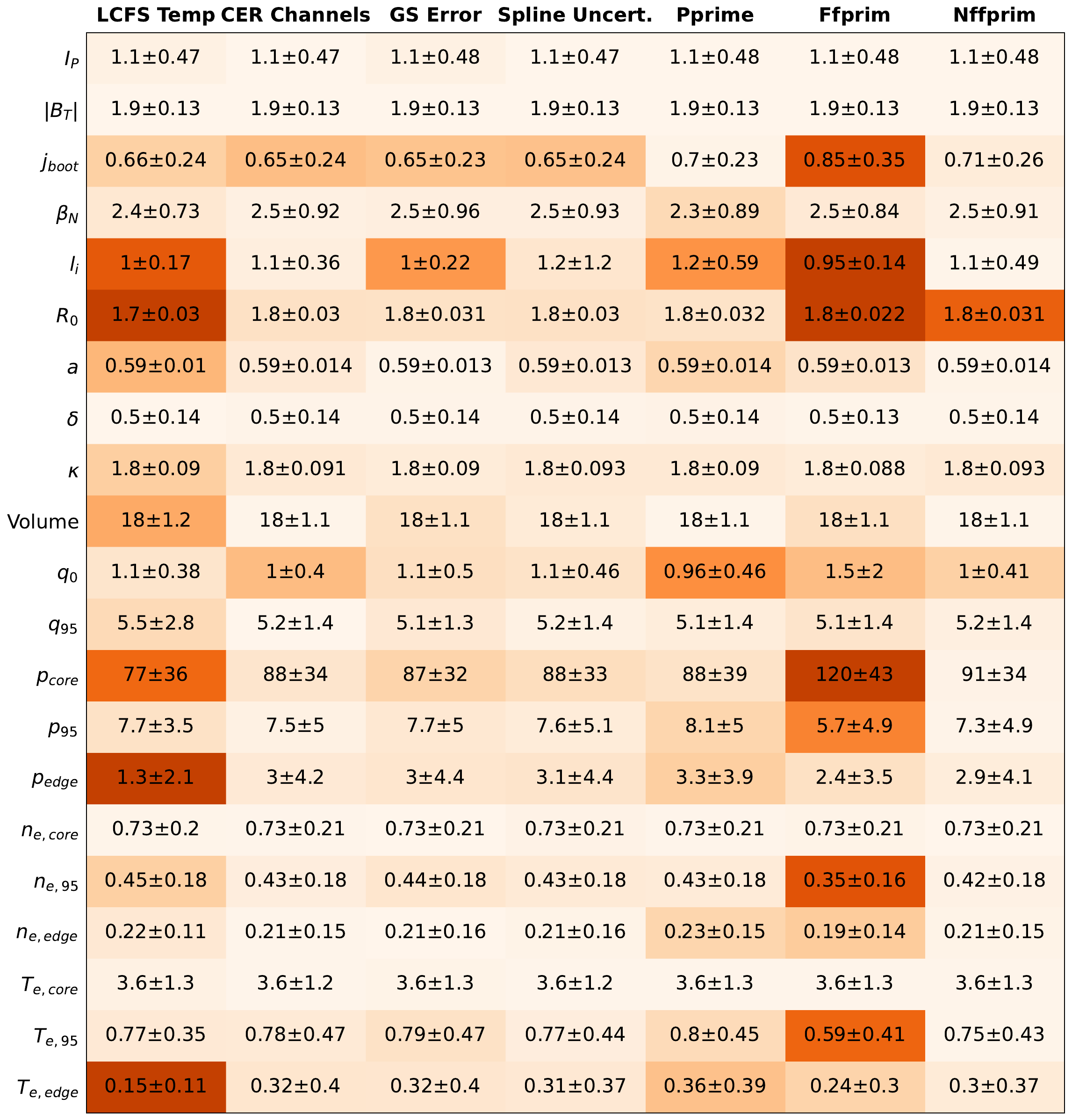}
    \caption{Parameter values of each CAKE variant compared to the default CAKE equilibria where each column has between 97-99 total equilbria. The intensity of color for each box is scaled by $(Mean_i -Row_{i,avg})/STD_i$: the mean value in a given box $i$ minus the average of the full row divided by the standard deviation of parameter $i$. This scaling will make variants with parameters that vary significantly from the base case more intense. }
    \label{fig:cakeVarTable}
\end{figure*}

The CAKE workflow settings were then adjusted for each setting listed in Section \ref{sec:cakeVar}, and scalar comparison statistics were compiled and are presented in Figure \ref{fig:cakeVarTable}. For these values, we report exact magnitudes instead of percent differences, where each box gives the mean of the CAKE variant compared to the default CAKE equilibria result plus or minus the standard deviation. With this metric, any parameter that is more than $1$ standard deviation away from the mean of each row is of interest, and so we color the intensity of each box by the deviation from the mean value of the row divided by the standard deviation of that parameter and CAKE variation. 

A first important CAKE setting is the electron temperature, as seen by the multiple dark colored boxes in the first column. Expected results of this are the large effects on $T_{e,edge}$ and $p_{edge}$, but there are smaller, yet noticeable effects on $p_{core}$ and $l_i$. Additionally, $R_0$ sees a significant shift, most likely to be attributed to how the Thomson electron temperature measurements are shifted to align the separatrix in the reconstruction iteration process. 

Another clear result is that the location of $FF^\prime$ spline smoothing knots are incredibly important and cause large deviations in a number of equilibrium parameters, including $j_{boot}$, $R_0$, $l_i$, core pressure, and pedestal electron density and temperature. Zooming in to $R_0$, this emphasis on deviation may be caused by the very small standard deviation. $l_i$ also appears to be changed by the additional GS error setting, as well as the $P^\prime$ smoothing spline locations. $P^\prime$ smoothing spline locations are also seen as higher impact parameters, as that column is generally higher intensity than the other variant columns. 

\section{Equilibria Reconstruction Effect on MHD stability}\label{sec:MHD}

We test the impacts of variations between the Manual dataset and CAKE dataset with results of MHD stability codes DCON \cite{glasser_direct_2016} and STRIDE \cite{glasser_robust_2018,glasser_riccati_2018}. DCON provides a single value of $\delta W$ indicating the ideal kink stability where $\delta W>0$ is stable and $\delta W<0$ is unstable as there is free energy in the equilibria perturbation. STRIDE provides the classical tearing mode stability metric $\Delta^\prime$, where we report $\Delta^\prime_{22}$ for the $m/n=2/1$ rational surface for poloidal mode number $m$ and toroidal mode number $n$.  For $\Delta^\prime$, there is a stabilizing effect from this term in the modified Rutherford equation~\cite{carrera_island_1986} for $\Delta^\prime<0$ and a destabilizing effect for $\Delta^\prime>0$. However, for full TM stability we would need a method of calculating $\Delta^\prime_\textrm{crit}$ where $\Delta^\prime>\Delta^\prime_\textrm{crit}$ represents an unstable equilibrium and $\Delta^\prime<\Delta^\prime_\textrm{crit}$ is a stable equilibrium. We will begin with a case study of a few equilibria that we vary with the OMFIT module PROcreate \cite{meneghini_integrated_2013,meneghini_integrated_2015} to approximate how numerically stable DCON is to forcing changes to equilibrium parameters. Then, we will leverage our full database to see how frequently the Manual versus CAKE equilibria disagree on both ideal kink and classical tearing stability. 

\subsection{Case study}

\begin{figure*}
    \centering
    \includegraphics[width=0.6\linewidth]{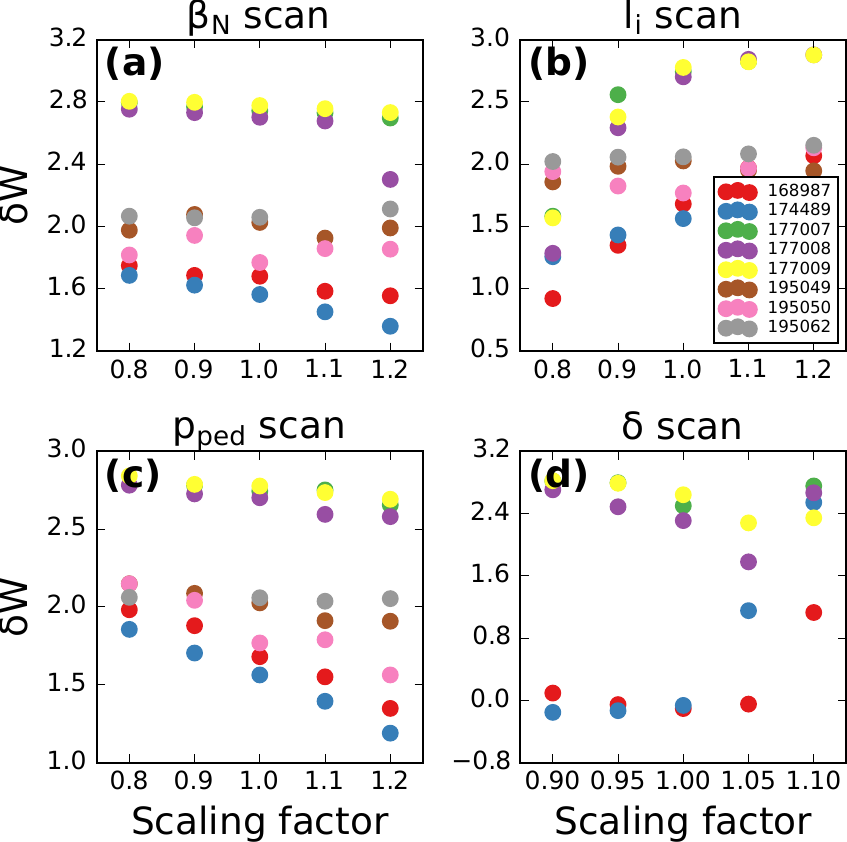}
    \caption{PROcreate scans of \textbf{(a)} $\beta_N$, \textbf{(b)} $l_i$, \textbf{(c)} pedestal pressure, and \textbf{(d)} triangularity ($\delta$) for the shot numbers given in the legend of \textbf{(b)}. The horizontal axis of each scan represents the scalar multiplied by the given parameter to recreate the equilibrium in PROcreate. Note that in \textbf{(d)} triangularity is only scanned in the range of $0.9$ to $1.1$ while the rest are scanned across a larger range of $0.8$ to $1.2$. }
    \label{fig:PCscans}
\end{figure*}

When comparing MHD stability of equilibria, the first step is defining a baseline to compare changes in stability parameters to. To set this baseline, we can artificially manipulate equilibria to change stability-relevant parameters, e.g. $\beta_N$ and $l_i$. If the equilibrium reconstruction process and the MHD stability code are robust, changes in $\delta W$ caused by changing relevant equilibrium quantities should be greater than a difference in $\delta W$ caused by changing the reconstruction method. 

We model this study in part on a previous $\delta W$ stability study~\cite{hanson_resistive_2021}, finding good agreement with those results. We expand our analysis to use $8$ base equilibria to avoid potential outlier cases that may see no change to $\delta W$. In Figure \ref{fig:PCscans}, we modify our reference CAKE equilibria to change $\beta_N$, $l_i$, pedestal pressure, and triangularity to quantify how much physically relevant changes to equilibria affect the calculated $\delta W$ values. General trends agree well with previous results~\cite{hanson_resistive_2021}, and we observe that increases in $\beta_N$ lead to lower $\delta W$ while higher values of $l_i$ lead to higher values of $\delta W$. While there is no full transport model taking into account all effects of changes to plasma shape to equilibrium profiles, small changes to triangularity and the corresponding effects to $\delta W$ in Figure \ref{fig:PCscans}.d represent some of the larger changes in $\delta W$ value. 

We estimate the effect of scanning these physical quantities on the $\delta W$ calculation. From the $\beta_N$ scans we see $\delta W$ changes by around a maximum of $0.5$, from $l_i$ changes can approach $1$, from $p_{ped}$ changes fall around $1$, and from triangularity scans changes are up to $2$. These scans are changing physical quantities of the plasma, so changes to $\delta W$ are reasonable. Consequently, we should be skeptical of an equilibrium pair of Manual and CAKE equilibria that are supposed to be modeling the same plasma that find $\delta W$ values that differ by more than $1$ or even $0.5$.

\subsection{DCON Stability Comparison}
\begin{figure*}
    \centering
    \includegraphics[width=0.7\linewidth]{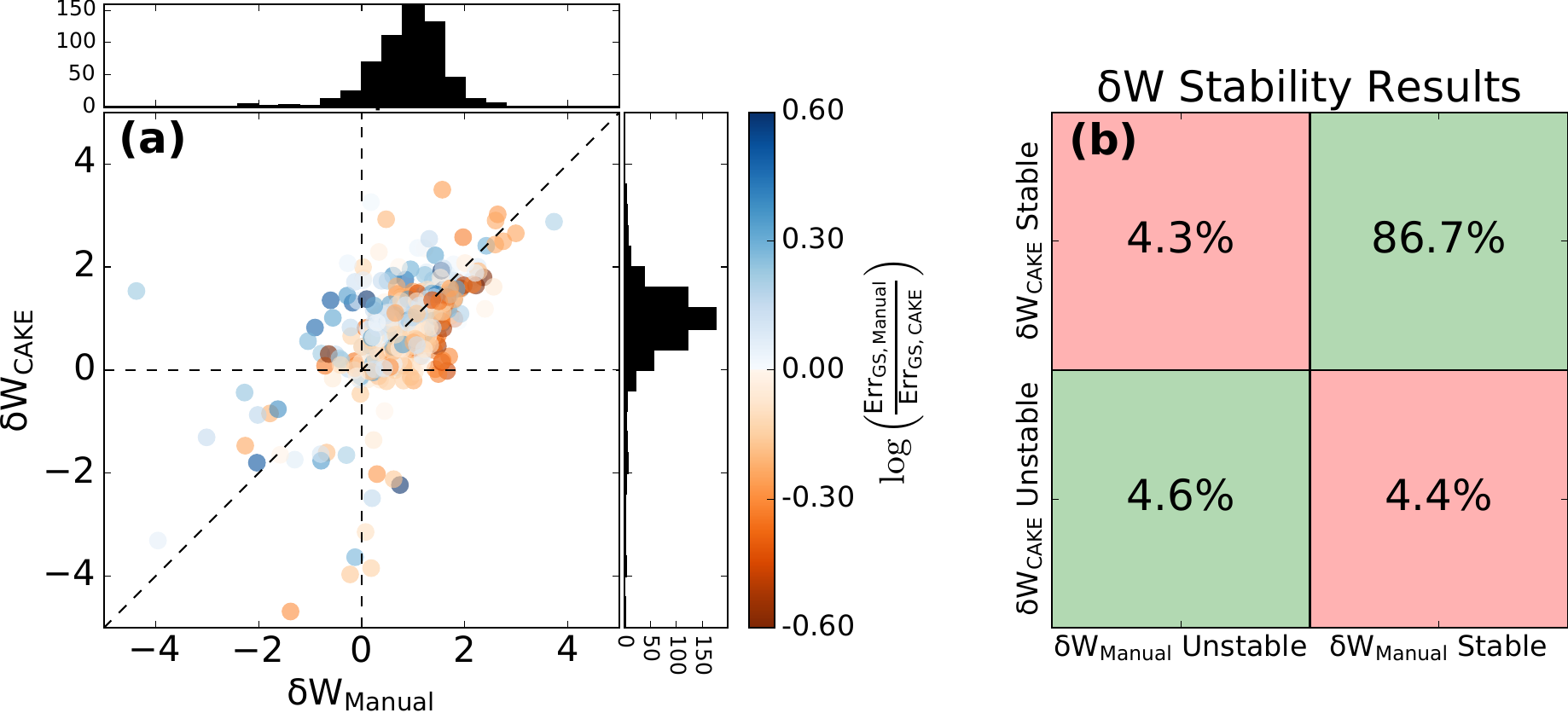}
    \caption{Database comparison between DCON $\delta W$ stability results for $552$ pairs of Manual Expert and CAKE equilibria. \textbf{(a)} Manual Expert values are plotted along the horizontal axis and CAKE values are plotted along vertical axis. The logarithm of the GS error of each equilibria pair is used to color each point meaning all good equilibria will have negative GS errors. Equal errors between the equilibria pair represent white, blue represent an equilibria pair where the Manual equilibrium GS error is worse than the CAKE equilibrium error, and red represents CAKE GS error being worse than the Manual equilibrium error. \textbf{(b)} Percentages of changes in stability classification due to different equilibria where the percent of points in each quadrant is given. }
    \label{fig:fullDBDCON}
\end{figure*}

Next, we turn to the full database of Manual and CAKE equilibria, where Figure \ref{fig:fullDBDCON} shows the full database of DCON results. There were 1238 Manual equilibria DCON runs and 1034 CAKE equilibria DCON runs. This left $785$ remaining equilibrium pairs. Because the $\delta W$ value changes depending on the edge $\psi$ to integrate to in relation to rational surfaces, equilibrium pairs were removed if the rational surface of the edge integration limit was not within $0.5$ between the Manual and CAKE equilibria. Additionally, we filtered out the Manual equilibria with extremely poor GS error, leaving 552 equilibrium pairs. 

For each equilibrium, we use the poloidally integrated GS error which we then take the maximum of in flux coordinate space, designated $Err_{GS,Manual}$ and $Err_{GS,CAKE}$ for the Manual and CAKE equilibria, respectively. We then color each point in Figure \ref{fig:fullDBDCON}.a by the logarithm of the ratio of these errors. Because these errors are in the range of $10^{-3}$ to $10^{-1}$, we interpret the colors as: red represents an equilibrium pair where the Manual equilibrium has lower GS error than the CAKE equilibrium and blue represents an equilibrium pair where the CAKE GS error is lower than the Manual GS error.  

Figure \ref{fig:fullDBDCON} shows significant spread in $\delta W$ values. We see differences of more than the $0.5-1$ range we found previously in the artificial $\delta W$ scans. This shows that the effects of equilibria reconstruction can change $\delta W$ values on the same order of physically relevant changes to the plasma equilibrium. 

Next, we observe that most red points fall below the $\delta W_{Manual}=\delta W_{CAKE}$ line, while most blue points are above that line. This implies that lower GS error leads to higher $\delta W$ values, as a blue point represents lower $Err_{GS,CAKE}$ and being above the diagonal means $\delta W_{CAKE}>\delta W_{Manual}$. This supports intuition that GS error is material to MHD stability calculations. 

Finally, we look to Figure \ref{fig:fullDBDCON}.b, where we classify each equilibrium pair into one of four quadrants, showing the percentage of points in each region. Any point that falls in the top right quadrant represents $\delta W_{Manual}>0$ and $\delta W_{CAKE}>0$, indicating agreement that the plasma is stable. The bottom left quadrant represents both $\delta W<0$ and agreement that the plasma is unstable. The upper left and lower right quadrants indicate disagreement on the stability of the plasma. We find that approximately $91\%$, the sum of the two green regions, of equilibrium pairs are in agreement as either stable or unstable to ideal kink stability. 

\subsection{STRIDE Stability Comparison}
\begin{figure*}
    \centering
    \includegraphics[width=0.7\linewidth]{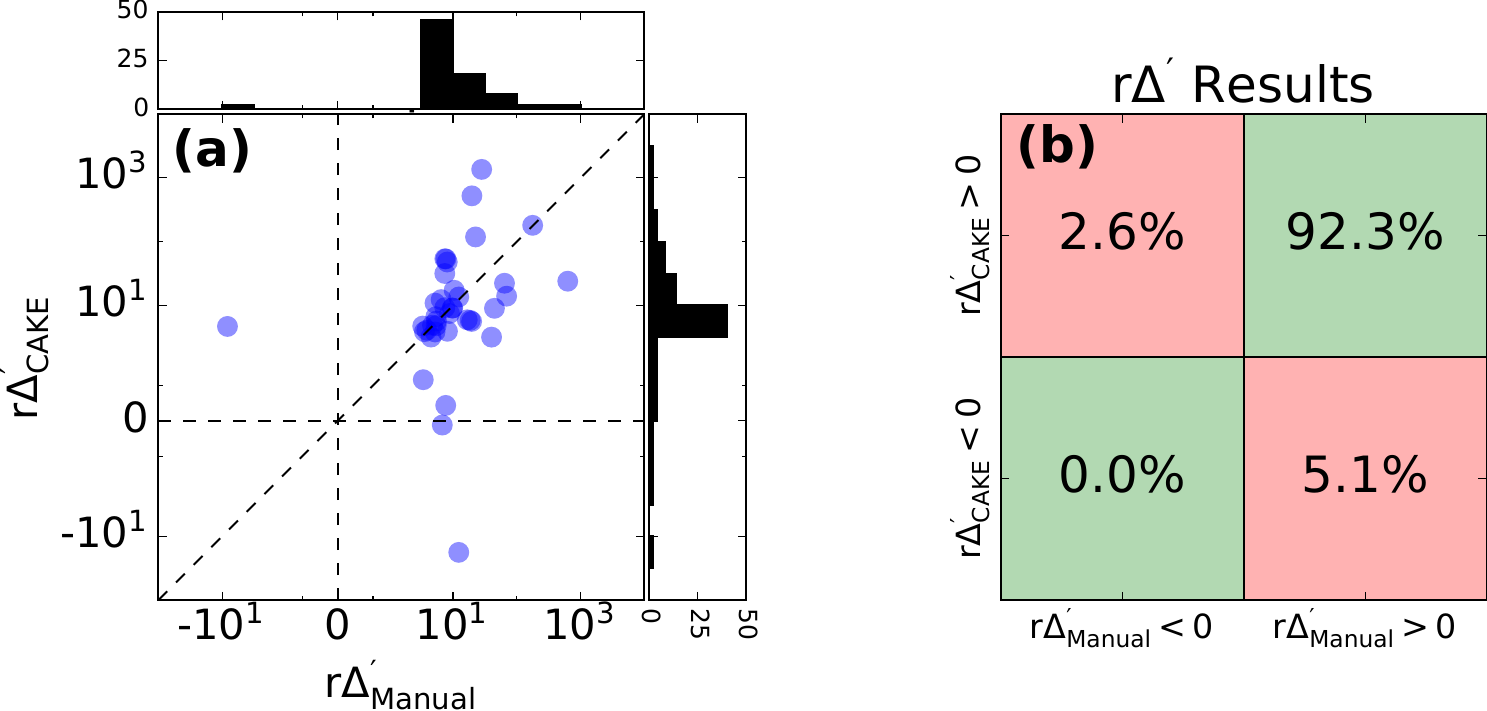}
    \caption{STRIDE runs for $78$ Manual and CAKE equilibria. \textbf{(a)} $\Delta^\prime$ tearing stability parameter plotted with symmetric log axes for each equilibrium pair. \textbf{(b)} classifies each point from \textbf{(a)} depending on whether $\Delta^\prime$ for each equilibria is stable or unstable, with the percent of pairs falling in each quadrant labeled. }
    \label{fig:DB_STRIDE}
\end{figure*}

We use the STRIDE\cite{glasser_robust_2018,glasser_computation_2016} classical tearing stability code to compare the numerical stability of our EFIT databases. We report values of the dimensionless quantity $r\Delta^\prime$, where $r$ is the minor radius of the $q=2$ surface, as we are concerned with the stability of $m,n=2/1$ TMs. Figure \ref{fig:DB_STRIDE} shows STRIDE results, including $1408$ runs on Manual equilibria and $266$ runs on CAKE equilibria. We filtered out equilibria where either $\delta W$ value was negative, equilibria where there were different numbers of rational surfaces, and equilibria without $q=2$ surfaces. This left only $78$ shared equilibria to compare. In Figure \ref{fig:DB_STRIDE}.a we see the differences in $\Delta^\prime$ values. With the symmetric log axes, there is extremely high variance in the magnitudes produced from the $r\Delta^\prime$ calculation where in some cases the values differing by more than a factor of $100$ for an equilibrium pair. This wide variance in $r\Delta^\prime$ values shows STRIDE is extremely sensitive to the equilibrium input, as expected by dependencies on the equilibrium profile gradients, and is likely not adequate for modified Rutherford equation modeling~\cite{urso_application_2010}, where the exact magnitude of $r\Delta^\prime$ is important for tearing mode growth rates.

Outside of just the magnitude difference, as we look to see stabilizing versus destabilizing effects in Figure \ref{fig:DB_STRIDE}.b, we see a slightly different picture. In the absence of a robust method to calculate the important $\Delta^\prime_\textrm{crit}$ parameter that would define tearing stability, we classify stabilizing versus destabilizing effects of $r\Delta^\prime$ and find that similarly to the DCON database results, the equilibrium pairings agree on tearing stability or instability for around $75\%$ of pairings. While there are fewer equilibria pairs than used in the DCON comparison, an agreed stability classification success rate of $75\%$ shows there is some consistency in $r\Delta^\prime$ stability results.

\section{Discussion, Conclusions, and Best Practices}\label{sec:conc}

We have presented a large database analysis comparing different equilibrium reconstruction methods and how their scalar and MHD stability parameters compare. The scalar parameter results agree well with expected results of shape parameters being robust, while parameters such as $q_0$ or $p_{edge}$, which are more reliant on plasma profiles or data-poor quantities, have higher uncertainties across the reconstruction methods. While these are similarly seen in the different setting variants to CAKE, we also find that specific changes to the equilibrium profiles (artificially introduced through modifications of the LCFS temperature or the splining locations in the $FF^\prime$ profile) result in large changes in the final equilibria, highlighting the importance of accurate, self-consistent kinetic profiles in equilibrium reconstructions. 

Looking at the downstream effects of variations in equilibrium reconstructions on MHD stability calculations, artificial physical parameter scans show changes in the ideal stability $\delta W$ around $0.5-1$, and when comparing reconstruction methods differences in $\delta W$ by this amount are suspect. Comparing $\delta W$ values across the Manual and CAKE databases, we find most of the results fall within this reasonable range. Importantly, we find that for reasonably good GS errors, the relative $\delta W$ value increases with relative GS error. $15\%$ of all equilibrium pairs included in this study had at least one $\delta W<0$. However, all of these plasmas were stable enough to produce kinetic equilibria and could not have been ideally unstable. This discrepancy could be a result of poor reconstructions leading to poor $\delta W$ values, unaccounted for wall stabilizing effects providing additional stability, or known issues of $\delta W<0$ calculations in advanced tokamak plasmas.

The $\Delta^\prime$ calculation from STRIDE is more numerically unstable and can change by factors of $10$ or $100$. These large deviations in $\Delta^\prime$ would create problems for tearing mode analysis as part of the modified Rutherford equation. The dependence on profile gradients is the likely culprit as both the $P^\prime$ and $FF^\prime$ profiles can vary greatly with equilibrium reconstruction method. 

In all, this work documents how variations in equilibrium reconstruction techniques and uncertainties in equilibrium reconstructions can impact the perceived results of basic tokamak analysis and conclusions drawn from more advanced physics models that build upon these reconstructions. While this represents a first attempt at validation of large scale database analysis of equilibrium reconstruction pipelines, it also highlights the importance of uncertainty quantification in tokamak research that focuses on a small subset (or individual) plasma scenario. We note that all of these tools and automated workflows are constantly under improvement and, as such, the numbers provided here should not be seen as absolute, but rather as a baseline upon which to improve consistency in how equilibria are reconstructed. There is significant variation in many parameters, especially in quantities such as $j_{boot}$ and $p_\mathrm{ped}$, which are both crucial in many downstream physics analyses. Based on these results, we advocate that more tokamak data analyses should include multiple equilibria for a single plasma or other explicit methods to quantify uncertainty. In particular, important profile quantities, like pedestal temperatures and densities, can be affected by choices made during the equilibrium reconstruction process, which may have unintended effects on other quantities in the plasma or on detailed conclusions reached through more advance physics tools.

\section*{Acknowledgment}

This material is based upon work supported by the U.S. Department of Energy, Office of Science, Office of Fusion Energy Sciences, using the DIII-D National Fusion Facility, a DOE Office of Science user facility, under Award DE-FC02-04ER54698. Additionally, this material is supported by the National Science Foundation Graduate Research Fellowship under Grant No. DGE-2039656 and by the U.S. Department of Energy, under Awards DE-SC0015480. 

\section*{Disclaimer}

This report was prepared as an account of work sponsored by an agency of the United States Government. Neither the United States Government nor any agency thereof, nor any of their employees, makes any warranty, express or implied, or assumes any legal liability or responsibility for the accuracy, completeness, or usefulness of any information, apparatus, product, or process disclosed, or represents that its use would not infringe privately owned rights. Reference herein to any specific commercial product, process, or service by trade name, trademark, manufacturer, or otherwise does not necessarily constitute or imply its endorsement, recommendation, or favoring by the United States Government or any agency thereof. The views and opinions of authors expressed herein do not necessarily state or reflect those of the United States Government or any agency thereof.

\section*{References}
\bibliographystyle{unsrt}
\bibliography{CAKE-DB.bib}% common bib file

\begin{thebibliography}{10}

\bibitem{buttery_advanced_2021}
R.J. Buttery, J.M. Park, J.T. McClenaghan, D.~Weisberg, J.~Canik, J.~Ferron,
  A.~Garofalo, C.T. Holcomb, J.~Leuer, and P.B. Snyder.
\newblock The advanced tokamak path to a compact net electric fusion pilot
  plant.
\newblock {\em Nuclear Fusion}, 61(4):046028, April 2021.

\bibitem{bourdelle_integrated_2025}
C~Bourdelle.
\newblock Integrated modelling of tokamak plasmas: progress and challenges
  towards {ITER} operation and reactor design.
\newblock {\em Plasma Physics and Controlled Fusion}, 67(4):043001, April 2025.

\bibitem{hassanein_potential_2021}
A.~Hassanein and V.~Sizyuk.
\newblock Potential design problems for {ITER} fusion device.
\newblock {\em Scientific Reports}, 11(1):2069, January 2021.

\bibitem{petty_diii-d_2019}
C.~C. Petty and {The DIII-D Team}.
\newblock {DIII}-{D} research towards establishing the scientific basis for
  future fusion reactors.
\newblock {\em Nuclear Fusion}, 59(11):112002, June 2019.
\newblock Publisher: Institute of Physics Publishing.

\bibitem{zohm_assessment_2013}
Hartmut Zohm.
\newblock Assessment of {DEMO} challenges in technology and physics.
\newblock {\em Fusion Engineering and Design}, 88(6-8):428--433, October 2013.

\bibitem{boivin_diii-d_2005}
R.~L. Boivin, J.~L. Luxon, M.~E. Austin, N.~H. Brooks, K.~H. Burrell, E.~J.
  Doyle, M.~E. Fenstermacher, D.~S. Gray, M.~Groth, C.-L. Hsieh, R.~J.
  Jayakumar, C.~J. Lasnier, A.~W. Leonard, G.~R. McKee, R.~A. Moyer, T.~L.
  Rhodes, J.~C. Rost, D.~L. Rudakov, M.~J. Schaffer, E.~J. Strait, D.~M.
  Thomas, M.~Van~Zeeland, J.~G. Watkins, G.~W. Watson, and C.~P.~C. Wong.
\newblock {DIII}-{D} {Diagnostic} {Systems}.
\newblock {\em Fusion Science and Technology}, 48(2):834--851, October 2005.

\bibitem{blum_problems_1990}
J.~Blum, E.~Lazzaro, J.~O'Rourke, B.~Keegan, and Y.~Stephan.
\newblock Problems and methods of self-consistent reconstruction of tokamak
  equilibrium profiles from magnetic and polarimetric measurements.
\newblock {\em Nuclear Fusion}, 30(8):1475, 1990.
\newblock Publisher: IOP Publishing.

\bibitem{belli_limitations_2014}
E~A Belli, J~Candy, O~Meneghini, and T~H Osborne.
\newblock Limitations of bootstrap current models.
\newblock {\em Plasma Phys. Control. Fusion}, 56(4):45006, 2014.

\bibitem{xing_cake_2021}
Z.A. Xing, D.~Eldon, A.O. Nelson, M.A. Roelofs, W.J. Eggert, O.~Izacard, A.S.
  Glasser, N.C. Logan, O.~Meneghini, S.P. Smith, R.~Nazikian, and E.~Kolemen.
\newblock {CAKE}: {Consistent} {Automatic} {Kinetic} {Equilibrium}
  reconstruction.
\newblock {\em Fusion Engineering and Design}, 163:112163, 2021.
\newblock tex.ids= xing\_cake\_2020.

\bibitem{sun_impact_2024}
Xuan Sun, Cihan Akcay, Torrin Bechtel, Scott Kruger, Lang~L Lao, Yueqiang Liu,
  Sandeep Madireddy, and Joseph McClenaghan.
\newblock Impact of various {DIII}-{D} diagnostics on the accuracy of neural
  network surrogates for kinetic {EFIT} reconstructions.
\newblock {\em Nuclear Fusion}, 64(8):086065, July 2024.

\bibitem{bechtel_accelerated_2022}
T.~A. Bechtel, A.~O. Nelson, L.~L. Lao, Z.~A. Xing, S.~P. Smith, R.~Nazikian,
  S.~Flanagan, D.~Schissel, L.~Stephey, R.~Thomas, S.~Williams, O.~Antepara,
  E.~Dart, E.~Koleman, and W.~Tang.
\newblock Accelerated {Workflow} for {Advanced} {Kinetic} {Equilibria}.
\newblock In {\em 2022 {First} {Combined} {International} {Workshop} on
  {Interactive} {Urgent} {Supercomputing} ({CIW}-{IUS})}, pages 1--5, Dallas,
  TX, USA, November 2022. IEEE.

\bibitem{denk_simultaneous_2025}
S~S Denk, T~B Amara, T~Odstrčil, L~Stagner, C~Akcay, T~Slendebroek, S~P Smith,
  M~A~Van Zeeland, T~Akiyama, and R~Nazikian.
\newblock Simultaneous kinetic profile and magnetic equilibrium inference with
  {Bayesian} integrated data analysis in preparation for {ITER}.
\newblock {\em Plasma Physics and Controlled Fusion}, 67(5):055014, May 2025.

\bibitem{lao_application_2022}
L~L Lao, S~Kruger, C~Akcay, P~Balaprakash, T~A Bechtel, E~Howell, J~Koo,
  J~Leddy, M~Leinhauser, Y~Q Liu, S~Madireddy, J~McClenaghan, D~Orozco,
  A~Pankin, D~Schissel, S~Smith, X~Sun, and S~Williams.
\newblock Application of machine learning and artificial intelligence to extend
  {EFIT} equilibrium reconstruction.
\newblock {\em Plasma Physics and Controlled Fusion}, 64(7):074001, July 2022.

\bibitem{lao_mhd_2005}
L.~L. Lao, H.~E. St~John, Q.~Peng, J.~R. Ferron, E.~J. Strait, T.~S. Taylor,
  W.~H. Meyer, C.~Zhang, and K.~I. You.
\newblock {MHD} equilibrium reconstruction in the {DIII}-{D} tokamak.
\newblock {\em Fusion Science and Technology}, 48(2):968--977, 2005.

\bibitem{logan_omfit_2018}
N~C Logan, B~A Grierson, S~R Haskey, S~P Smith, O~Meneghini, and D~Eldon.
\newblock \{{OMFIT}\} {Tokamak} {Profile} {Data} {Fitting} and {Physics}
  {Analysis}.
\newblock {\em Fusion Science and Technology}, 74(1-2):125--134, 2018.
\newblock Publisher: Taylor \& Francis.

\bibitem{madireddy_efit-prime_2024}
S.~Madireddy, C.~Akçay, S.~E. Kruger, T.~Bechtel Amara, X.~Sun,
  J.~McClenaghan, J.~Koo, A.~Samaddar, Y.~Liu, P.~Balaprakash, and L.~L. Lao.
\newblock {EFIT}-{Prime}: {Probabilistic} and physics-constrained reduced-order
  neural network model for equilibrium reconstruction in {DIII}-{D}.
\newblock {\em Physics of Plasmas}, 31(9):092505, September 2024.

\bibitem{poli_experimental_2016}
F~M Poli, P~T Bonoli, M~Chilenski, R~Mumgaard, S~Shiraiwa, G~M Wallace,
  R~Andre, L~Delgado-Aparicio, S~Scott, J~R Wilson, R~W Harvey, Yu~V Petrov,
  M~Reinke, I~Faust, R~Granetz, J~Hughes, and J~Rice.
\newblock Experimental and modeling uncertainties in the validation of lower
  hybrid current drive.
\newblock {\em Plasma Physics and Controlled Fusion}, 58(9):095001, September
  2016.

\bibitem{rodrigues_sensitivity_2016}
P.~Rodrigues, A.C.A. Figueiredo, D.~Borba, R.~Coelho, L.~Fazendeiro,
  J.~Ferreira, N.F. Loureiro, F.~Nabais, S.D. Pinches, A.R. Polevoi, and S.E.
  Sharapov.
\newblock Sensitivity of alpha-particle-driven {Alfvén} eigenmodes to
  q-profile variation in {ITER} scenarios.
\newblock {\em Nuclear Fusion}, 56(11):112006, November 2016.

\bibitem{abbate_large-database_2024}
J.~Abbate, E.~Fable, B.~Grierson, A.~Pankin, G.~Tardini, and E.~Kolemen.
\newblock Large-database cross-verification and validation of tokamak transport
  models using baselines for comparison.
\newblock {\em Physics of Plasmas}, 31(4):042506, April 2024.

\bibitem{snyder_edge_2002}
P.~B. Snyder, H.~R. Wilson, J.~R. Ferron, L.~L. Lao, A.~W. Leonard, T.~H.
  Osborne, A.~D. Turnbull, D.~Mossessian, M.~Murakami, and X.~Q. Xu.
\newblock Edge localized modes and the pedestal: {A} model based on coupled
  peeling–ballooning modes.
\newblock {\em Physics of Plasmas}, 9(5):2037--2043, May 2002.

\bibitem{nelson_time-dependent_2021}
Andrew~Oakleigh Nelson, Florian~M Laggner, Ahmed Diallo, David Smith, Z~Anthony
  Xing, Ricardo Shousha, and Egemen Kolemen.
\newblock Time-dependent experimental identification of inter-{ELM}
  microtearing modes in the tokamak edge on {DIII}-{D}.
\newblock {\em Nuclear Fusion}, 61(11):116038, 2021.
\newblock Publisher: IOP Publishing.

\bibitem{hassan_identifying_2022}
Ehab Hassan, D.R. Hatch, M.R. Halfmoon, M.~Curie, M.T. Kotchenreuther, S.M.
  Mahajan, G.~Merlo, R.J. Groebner, A.O. Nelson, and A.~Diallo.
\newblock Identifying the microtearing modes in the pedestal of {DIII}-{D}
  {H}-modes using gyrokinetic simulations.
\newblock {\em Nuclear Fusion}, 62(2):026008, February 2022.

\bibitem{nelson_robust_2023}
A.O. Nelson, L.~Schmitz, C.~Paz-Soldan, K.E. Thome, T.B. Cote, N.~Leuthold,
  F.~Scotti, M.E. Austin, A.~Hyatt, and T.~Osborne.
\newblock Robust {Avoidance} of {Edge}-{Localized} {Modes} alongside {Gradient}
  {Formation} in the {Negative} {Triangularity} {Tokamak} {Edge}.
\newblock {\em Physical Review Letters}, 131(19):195101, November 2023.

\bibitem{turco_causes_2018}
F.~Turco, T.C. Luce, W.~Solomon, G.~Jackson, G.A. Navratil, and J.M. Hanson.
\newblock The causes of the disruptive tearing instabilities of the {ITER}
  {Baseline} {Scenario} in {DIII}-{D}.
\newblock {\em Nuclear Fusion}, 58(10):106043, October 2018.

\bibitem{bardoczi_onset_2023}
L.~Bardóczi, N.J. Richner, and N.C. Logan.
\newblock The onset distribution of rotating m,n=2,1 tearing modes and its
  consequences on the stability of high-confinement-mode plasmas in {DIII}-{D}.
\newblock {\em Nuclear Fusion}, 63(12):126052, December 2023.

\bibitem{groebner_progress_2001}
R.J Groebner, D.R Baker, K.H Burrell, T.N Carlstrom, J.R Ferron, P~Gohil, L.L
  Lao, T.H Osborne, D.M Thomas, W.P West, J.A Boedo, R.A Moyer, G.R McKee, R.D
  Deranian, E.J Doyle, C.L Rettig, T.L Rhodes, and J.C Rost.
\newblock Progress in quantifying the edge physics of the {H} mode regime in
  {DIII}-{D}.
\newblock {\em Nuclear Fusion}, 41(12):1789--1802, 2001.
\newblock tex.ids= groebner\_progress\_2001.

\bibitem{ding_high-density_2024}
S.~Ding, A.~M. Garofalo, H.~Q. Wang, D.~B. Weisberg, Z.~Y. Li, X.~Jian,
  D.~Eldon, B.~S. Victor, A.~Marinoni, Q.~M. Hu, I.~S. Carvalho, T.~Odstrčil,
  L.~Wang, A.~W. Hyatt, T.~H. Osborne, X.~Z. Gong, J.~P. Qian, J.~Huang,
  J.~McClenaghan, C.~T. Holcomb, and J.~M. Hanson.
\newblock A high-density and high-confinement tokamak plasma regime for fusion
  energy.
\newblock {\em Nature}, 629(8012):555--560, April 2024.

\bibitem{avdeeva_accuracy_2024}
Galina Avdeeva, Kathreen~E Thome, John~W Berkery, Stanley~M Kaye, Joseph
  McClenaghan, O~Meneghini, Tomas Odstrcil, Steve~A Sabbagh, Sterling~P Smith,
  and Alan~D Turnbull.
\newblock Accuracy of kinetic equilibrium reconstruction of {NSTX} and
  {NSTX}-{U} plasmas and its impact on the transport and stability analysis.
\newblock {\em Plasma Physics and Controlled Fusion}, 66(11):115003, September
  2024.

\bibitem{osborne_enhanced_2015}
T~H Osborne, G~Jackson, Z~Yan, R~Maingi, D~Mansfield, B~Grierson, C~Chrobak,
  A~McLean, S~Allen, D~Battaglia, A~Briesemeister, M~Fenstermacher, G~McKee,
  and P~Snyder.
\newblock Enhanced {H}-mode pedestals with lithium injection in {DIII}-{D}.
\newblock {\em Nucl. Fusion}, 55(6):63018, 2015.

\bibitem{lao_reconstruction_1985}
L.~L. Lao, H.~St John, R.~D. Stambaugh, A.~G. Kellman, and W.~Pfeiffer.
\newblock Reconstruction of current profile parameters and plasma shapes in
  tokamaks.
\newblock {\em Nuclear Fusion}, 25(11):1611--1622, November 1985.
\newblock Publisher: IOP Publishing.

\bibitem{pfeiffer_onetwo_nodate}
W~W Pfeiffer, R~H Davidson, R~L Miller, and R~E Waltz.
\newblock {ONETWO}: {A} {COMPUTER} {CODE} {FOR} {MODELING} {PLASMA} {TRANSPORT}
  {IN} {TOKAMAKS}.

\bibitem{meneghini_integrated_2015}
O.~Meneghini, S.P. Smith, L.L. Lao, O.~Izacard, Q.~Ren, J.M. Park, J.~Candy,
  Z.~Wang, C.J. Luna, V.A. Izzo, B.A. Grierson, P.B. Snyder, C.~Holland,
  J.~Penna, G.~Lu, P.~Raum, A.~McCubbin, D.M. Orlov, E.A. Belli, N.M. Ferraro,
  R.~Prater, T.H. Osborne, A.D. Turnbull, and G.M. Staebler.
\newblock Integrated modeling applications for tokamak experiments with
  {OMFIT}.
\newblock {\em Nuclear Fusion}, 55(8):083008, August 2015.

\bibitem{meneghini_integrated_2013}
Orso Meneghini and Lang Lao.
\newblock Integrated {Modeling} of {Tokamak} {Experiments} with {OMFIT}.
\newblock {\em Plasma and Fusion Research}, 8(0):2403009--2403009, 2013.

\bibitem{glasser_direct_2016}
A~H Glasser.
\newblock The direct criterion of {Newcomb} for the ideal {MHD} stability of an
  axisymmetric toroidal plasma.
\newblock {\em Phys. Plasmas}, 23(7):72505, 2016.

\bibitem{glasser_robust_2018}
Alexander~S. Glasser and Egemen Kolemen.
\newblock A {Robust} {Solution} for the {Resistive} {MHD} {Toroidal}
  \${\textbackslash}{Delta}{\textasciicircum}{\textbackslash}prime\$ {Matrix}
  in {Near} {Real}-{Time}.
\newblock {\em Physics of Plasmas}, 25(8):082502, August 2018.

\bibitem{glasser_riccati_2018}
Alexander~S. Glasser, Egemen Kolemen, and A.~H. Glasser.
\newblock A {Riccati} solution for the ideal {MHD} plasma response with
  applications to real-time stability control.
\newblock {\em Physics of Plasmas}, 25(3):032507, March 2018.

\bibitem{carrera_island_1986}
R.~Carrera, R.~D. Hazeltine, and M.~Kotschenreuther.
\newblock Island bootstrap current modification of the nonlinear dynamics of
  the tearing mode.
\newblock {\em Physics of Fluids}, 29(4):899, 1986.
\newblock Publisher: AIP Publishing.

\bibitem{hanson_resistive_2021}
J.~M. Hanson, F.~Turco, T.~C. Luce, G.~A. Navratil, and E.~J. Strait.
\newblock Resistive contributions to the stability of {DIII}-{D} {ITER}
  baseline demonstration discharges.
\newblock {\em Physics of Plasmas}, 28(4):042502, April 2021.

\bibitem{glasser_computation_2016}
A.~H. Glasser, Z.~R. Wang, and J.-K. Park.
\newblock Computation of resistive instabilities by matched asymptotic
  expansions.
\newblock {\em Physics of Plasmas}, 23(11):112506, November 2016.

\bibitem{urso_application_2010}
L~Urso, R~Fischer, A~Isayama, {ASDEX Upgrade Team}, and {JT-60 Team}.
\newblock Application of the {Bayesian} analysis to the modified {Rutherford}
  equation for {NTM} stabilization.
\newblock {\em Plasma Physics and Controlled Fusion}, 52(5):055012, May 2010.

\end{thebibliography}

\end{document}